# Unveiling the Atomic and Electronic Structure at the Surface of the Parent Pnictide $SrFe_2As_2$

F. C. Niestemski[1], Von Braun Nascimento[2], Biao Hu[2], Ward Plummer[2], J. Gillett[3], Suchitra Sebastian[3], Ziqiang Wang[1], V. Madhavan[1]

**The parent compounds of the recently discovered iron-arsenic (pnictide) high temperature superconductors[1] transition into an intriguing spin density wave (SDW) phase at low temperatures. Progress in understanding this SDW state has been complicated by a complex band structure and by the fact that the spin, electronic, and structural degrees of freedom are closely intertwined in these compounds. Scanning tunneling microscopy (STM) measurements have added to this complexity by revealing different topographies with no consensus on the surface structure[2,3,4,5]. In this paper, we use a combination of high-resolution STM imaging and spectroscopy, and low energy electron diffraction (LEED) to determine the atomic and electronic structure of the parent pnictide $SrFe_2As_2$. Our data present a compelling picture of the existence of two coexisting homotopic structures on the surface. Based on this, we construct a simple model for the surface, which offers an explanation of the two classes of topographies seen by STM. STM spectroscopy shows that while the high energy density of states (DOS) profile is consistent with the Fe 3d and As 4p-electrons predicted by LDA[6,7], it is in better agreement with calculations that include electron correlations beyond LDA[8]. Importantly, we find a gap of ~15 meV in the low energy density of states on both structures which may be linked with the SDW or the observed surface reconstruction.**

[1] *Department of Physics, Boston College, Chestnut Hill, Massachusetts 02467, USA*
[2] *Louisiana State University, Department of Physics & Astronomy, Baton Rouge, LA 70803-4001, USA*
[3] *Quantum Matter Group, Cavendish Laboratory, University of Cambridge, Cambridge CB3 0HE, UK*

Understanding the interplay among the lattice, spin, charge, and orbital degrees of freedom in the parent pnictides is critical to addressing many fundamental physical properties in this material, including the mechanism of superconductivity. Most undoped pnictides exhibit a tetragonal to orthorhombic structural transition as well as a spin density wave (SDW) transition[9, 10, 11, 12] at low temperatures. The resultant state is metallic, albeit with low carrier density. Superconductivity arises either by applying pressure to the parent compounds or by chemical doping with extra carriers. In contrast to the high-$T_c$ cuprates, the pnictides are on the metallic side of the Mott transition, although it is actively debated whether magnetic order is driven primarily by itinerant electrons or localized spins[13, 14]. Optical spectroscopy data on $SrFe_2As_2$ (Sr122) show two gaps around 60 meV and 180 meV in the particle-hole excitations (2Δ) below the SDW transition temperature[15]. In contrast, angle resolved photoemission spectroscopy (ARPES) studies on the undoped compounds show contradictory evidence for gaps near the Fermi energy ($E_F$)[4, 16, 17, 18] and remain inconsistent [4, 16, 17, 18, 19, 20] among themselves and with quantum oscillation measurements[21] on many details of the Fermi surface topology. This raises questions about the nature of the surface structure its effect on the density of states.

STM is a proven technique to simultaneously map out the topography and the low energy electronic density of states (DOS) in complex materials such as the high-$T_c$ cuprates. Information from STM is particularly important in the pnictides since there is experimental and theoretical evidence that the electronic structure, lattice dynamics, and the SDW ordering are intertwined with the detailed geometry of the Fe-As plane. To date, there have been no systematic STM spectroscopic studies dedicated to the parent pnictides. Prior STM images of the 122 ($AFe_2As_2$, A = Sr, Ba) compounds show a striped topography (2 X1) with 8Å periodicity[2,3,4], which was attributed to a dimer reconstruction of atoms in the As or the Sr/Ba planes[2,4] but a clear chemical identification of the superstructure was not possible. On the other hand, a recent detailed LEED and STM study showed that the ordered surface of Ba122 is an As bulk-like termination with the 1x1 unit cell (*p2mm*) expected for a bulk orthorhombic structure with no structural distortion in the surface plane and the Ba atoms randomly distributed on the surface[5].

In this paper, we present low temperature (5K), high-resolution STM measurements on $SrFe_2As_2$ single crystal samples. Samples were cleaved both at room temperature (warm cleave) and at liquid nitrogen temperatures (cold cleave). In the following we will use the symmetry notation associated with the high temperature tetragonal phase to describe the surface unit cells. We observe two types of ordered surfaces, regions showing a square lattice with ~5.7 Å periodicity ($\sqrt{2}$ region) (Figs 1a, b) and striped regions with ~8Å periodicity across the stripes and ~4Å periodicity along the stripes (2 X 1 region) (Fig 1c). The stripes are found to be oriented 45° with respect to the lattice vectors of the $\sqrt{2}$ region (Fig. 1b) and observed in samples cleaved either warm or cold (Figs 1c and 2a). Both areas are peppered with bright clusters of various sizes as well as dark regions, which could be attributed to lattice defects in the exposed planes, subsurface, or to impurities atoms on the surface. There is a propensity for these defects and impurity clusters to be aligned along one of the high symmetry directions of the plane. While these two topographies have been observed by various STM groups, their occurrence on the same sample allows us to investigate the relationship between them using imaging and spectroscopy.

As a first step towards understanding the surface, we consider the crystal structure

(Fig. 2). In the low temperature orthorhombic phase both the Sr and As atoms are in an approximately square lattice of 4Å periodicity barring a small difference between the a- and b-axis (a = 3.9382 Å, b = 3.8975 Å)[22]. The new in-plane unit cell is ($\sqrt{2}$ x $\sqrt{2}$) R45$^0$ in the tetragonal notation or (1x1) in the orthorhombic notation. The layered nature of the Sr and Fe-As planes suggests that the cleave most likely occurs between the two Fe-As bilaygrs exposing Sr and/or As atoms. Since the Sr atoms are more weakly bound, they are expected to distribute statistically on the two halves of the cleaved crystal.

Figure 2 shows a schematic of the Sr122 surface. We illustrate the unreconstructed lattice by marking the location of each As/Sr ion in the ~4Å square lattice plane by a point (Fig. 2c). In general, the contrast and exact location of these lattice points in STM topography is influenced the charge density profile of the surface, which is coupled closely the structure, local density of states and the nature of the orbitals. The red dots/ovals in the schematic (Figs 2d, e) capture the observed contrast in the striped and $\sqrt{2}$ regions respectively. We begin with a discussion of the $\sqrt{2}$ region where half the atoms are invisible to the STM. The symmetry (*p2mm*) of this region provides important clues to its origin. In the low temperature orthorhombic phase, a combination of the spin structure and the lattice distortion results in a unit cell of As atoms with 5.67Å periodicity. Since this is the natural symmetry of the As layer, it is certainly conceivable that the $\sqrt{2}$ regions region represents the As plane. Furthermore, the chemical identity of these planes can be addressed by quantitative low energy electron diffraction (LEED). Similar to a recent study of Ba122[5], LEED intensity-voltage (*I(V)*) curves were quantitatively analyzed to determine the surface termination by minimizing the reliability factor (R-factor). These fits indicate that the 1 X1 ordered layer in Sr122 is also composed of As atoms (the Fe-As plane) (Fig. S3). Although Sr and As planes can be treated on the same footing for the STM data, based on LEED we identify the $\sqrt{2}$ region as the As plane.

The striped regions show dimerized rows of atoms (inset to fig 2b), a structure that quite different from the $\sqrt{2}$ region. Remarkably, a closer look at the $\sqrt{2}$ region as well as the boundary between the stripe and $\sqrt{2}$ regions provide evidence that the stripe and $\sqrt{2}$ regions are intimately connected. Large area STM images show the transition between these regions (Fig. 2g and S1). We find that the stripes emerge gradually from the $\sqrt{2}$ region with no step edge between them. This intimate relationship between these regions is further emphasized by one of the most intriguing properties of the $\sqrt{2}$ region, i.e., the formation of local short stripes. STM images show that whenever two impurities or defects are connected by a multiple of the stripe vector, short stripes (ladder-like features in the topography) appear (Fig. 1a and 2h). These local stripes have the same periodicity and orientation as the long-range stripes (Figs 1 and S2). In our model (Fig. 2f), the formation of these local stripes is triggered by new boundary conditions due to defects and impurities, or dislocations. As seen in Fig. 2h, this idea is clearly consistent with the STM topography. The instability towards local stripes in the $\sqrt{2}$ region reiterates the homotopic nature of the two surfaces and implies that the two orders are close in energy.

Having identified the striped and $\sqrt{2}$ region as two manifestations of the same plane, we now address the question of the driving force behind the stripe formation. Since materials with nested Fermi surfaces are often susceptible to a charge density wave (CDW) instability, which could in turn be associated with a lattice reconstruction, we consider the possibility of a CDW order on the surface that is associated with the SDW order in the bulk. It should be noted however, that the direction of the stripes are 45° rotated from that of the SDW in the Fe plane and the wave vector associated with the stripes does not correspond to any nesting vector of the observed or calculated Fermi surfaces [6, 14, 23, 24, 25, 26]. More significantly, from the *dI/dV* maps at various energies (Fig. 3), we observe that the stripe pattern does not show any contrast reversal (phase shift in the spatial pattern at a pair of positive and negative bias), which is considered a hallmark of CDW order in STM data[27]. We thus conclude that the observed structures are unlikely to arise from a conventional CDW. The other possible driving forces are either a relaxation on the surface of strain due to the underlying Sa/Ba/Ca layer or a freezing of a surface phonon mode. High resolution LEED and phonon measurements as well as theory calculations for the surface energetics will be important in determining the driving force for this reconstruction.

To examine the possible effects of the observed order on the local density of states (*LDOS*), we turn to STM spectroscopy (*dI/dV* measurements). STM spectra in regions free from impurities reveal a few common features in both regions. High energy spectra reveal a steep increase in the density of states (DOS) ~500 meV below the Fermi energy ($E_F$) (Fig. 4a). A similar rapid increase in the DOS is observed as we go ~200 meV above $E_F$. These features within an electron volt of the Fermi energy are in general agreement with the DOS contributions from both the As 4p-electrons and the Fe 3d-electrons predicted by the LDA[6, 7] but are in better agreement with calculations that the effects of correlations[8].

Zooming in on the medium energy *DOS* reveals two clear features at ~250 meV below $E_F$ and ~150 meV above $E_F$ (Figs 3i and 4a) with a perceived suppression in density of states below these energies. While it is tempting to associate these features with a gap of 2Δ ~ 400 meV arising from to a collective state (the SDW or a CDW for example), low temperature spectroscopy alone is not sufficient to identify this feature as such. A density of state pile up from the stripped phases superimposed on a highly non-linear background could mimic a gap-like feature. Moreover, this suppression might arise from the complex band structure and more data is necessary before we can positively identify the origin of this gap-like suppression. The feature ~+150 meV however shows an interesting response to the stripes. *dI/dV* maps in the striped regions reveal that the spectral weight of the stripes mainly lies within a narrow range of energies 153 meV +- 20 meV above $E_F$ (Fig. 3d). This indicates that this feature may be identified as a local increase in the density of states at this energy caused by the stripe formation. We note that the presence of this resonance does not preclude a gap since such a resonance might also occur at a gap edge. Remarkably, the local stripes in the $\sqrt{2}$ region also show an enhanced contrast around +150 meV once again confirming that they have the same origin as the long stripes (Fig. 3h).

At low energies we observe a ubiquitous gap like feature with 2Δ ~ 30-40 meV (fig 4b). The fact that this feature is pinned to the Fermi energy suggests a collective phenomenon. One possible explanation is that this gap arises from the structural transition at the surface. The other possibility is that it is a gap associated with the SDW phase[15]. The latter implies an itinerant SDW where the depletion of the DOS near Fermi level due to the opening of an SDW gap on parts of the Fermi surface results in a metallic SDW phase with low carrier density[15, 20]. Temperature dependent measurements are needed to unambiguously identify the origin of this gap feature.

Our STM studies of the parent Sr122 pnictides show striped and $\sqrt{2}$ regions in the same sample for the first time, allowing us to form a coherent picture of the origin of the two surfaces. Our data ties together the different topographies observed by other groups on the pnictides and reveals that the As surface itself has a propensity to form stripes. STM spectroscopy and maps of the two regions show that the density of states is similar over large energy scales, supporting our conclusion that the two surfaces are homotopically equivalent manifestations of the same As plane. The spectroscopy also reveals a low energy gap pinned to $E_F$, most likely a manifestation of the coupling of the different degrees of freedom in these materials, although further work is necessary to identify the origin of this gap. Finally, our spectroscopic studies of the parent compound form a basis of comparison for studies of the superconducting state in this material.

Acknowledgement: We acknowledge H ong Ding, E.W. Hudson, and A.V. Balatsky for valuable discussions.

Figure Captions

Figure 1. **STM Topography of striped and $\sqrt{2}$ regions a.** 256 Å x 256 Å STM image of $\sqrt{2}$ region of sample #1, bias voltage ($V_b$) = -160 mV, tunnel current (I) = 50 pA. Sample #1 was cleaved cold ~77K **b.** 50 Å x 50 Å image (low pass filtered to improve clarity of atoms) of a $\sqrt{2}$ region of sample #1 showing the relationship between the lattice in the $\sqrt{2}$ region and the direction of the stripes. **c.** 256 Å x 256 Å image of striped region in sample #1, $V_b$ = -100 mV, I= 200 pA. Stripes were observed only along two directions (two red arrows) in a given sample.

Figure 2. **Schematic of Sr122 surface a.** Unit cell of $SrFe_2As_2$ and top view showing the a-b planes. **b.** 45 Å X65 Å topograph of striped region obtain with junction resistance ($R_J$)= 55 MΩ. Inset is 45 Å X 12 Å ($R_J$ = 10 MΩ) showing that the stripes are actually dimerized rows of atoms. We note that in Sr122 it is necessary to image at rather low junction resistances to see the dimerization. **c.** Schematic model of the As/Sr layer with no reconstruction. The points are arranged in a square lattice 4 Å apart. The orthorhombic distortion has been ignored for simplicity. **d. and e.** Schematic of the observed contrast in the striped and $\sqrt{2}$ regions. In the $\sqrt{2}$ regions half the lattice points (black dots) are present but not visible to STM topography **f.** $\sqrt{2}$ region with impurities (blue dots) which cause change in the pattern resulting in local stripes as shown in **h. g.** 375 Å x 980 Å topography ($R_J$ = 0.8 GΩ) showing a transition from the striped to the $\sqrt{2}$ region. The lack of a step edge between the two regions implies that they are both the same plane. **h.** 96 Å x 96 Å zoom in of $\sqrt{2}$ region ($R_b$ = 1.7 GΩ). Local stripes with 8Å periodicity are created when two impurities can be connected by multiples of the stripe vector, as demonstrated in **f.**

Figure 3. **Spatial *dI/dV* maps and medium energy spectra a**. 256 Å x 256 Å topography and corresponding dI/dV maps (**b, c,** and **d**) at -150 mV, +20 mV, +150 mV respectively. ($R_J$ = 15.0 GΩ). **e.** 96 Å x 96 Å topography of the $\sqrt{2}$ area with local stripes ($R_J$ =7.3 GΩ) **f, g, h**. *dI/dV* maps of **e**. **i.** Series of *dI/dV* spectra perpendicular to the stripes at 1Å intervals, ($V_b$ =+400 mV, I= 300 pA) showing the modulation of the resonance ~ +150 mV in response to the stripes. Spectra are offset along x-axis for clarity.

Figure 4. **High and low energy spectra a.** High energy averaged *dI/dV* spectrum of region shown in **1a** showing the steep rise in density of states ~ 500 meV below and ~200 meV above $E_F$ as well as the medium energy features, the counterparts of fig. 3i now in the $\sqrt{2}$ region. **b.** Series of *dI/dV* spectra at different points of a striped region showing the ubiquitous low energy ~30 meV (2Δ) gap.

Figure 1

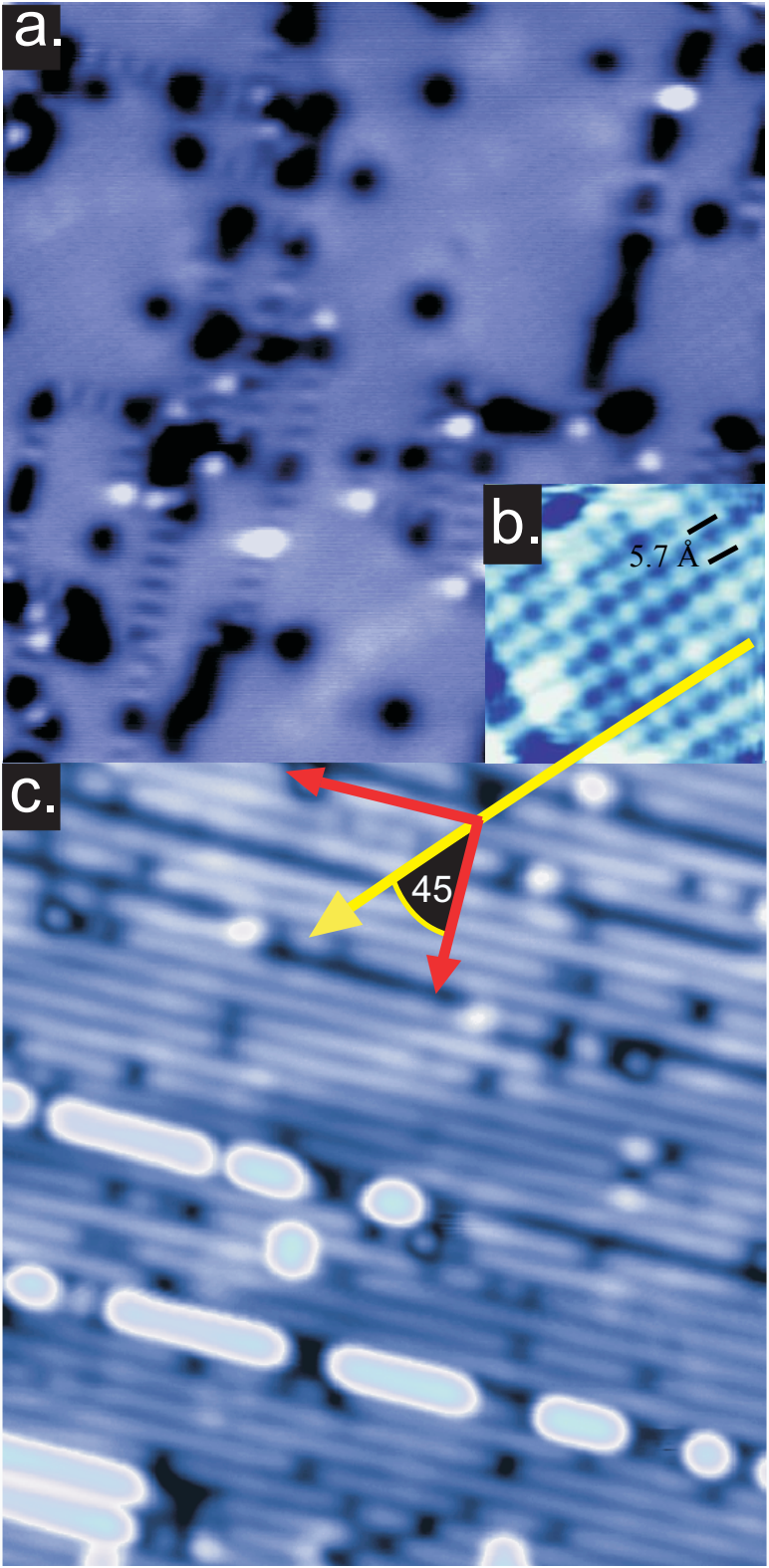

# Figure 2

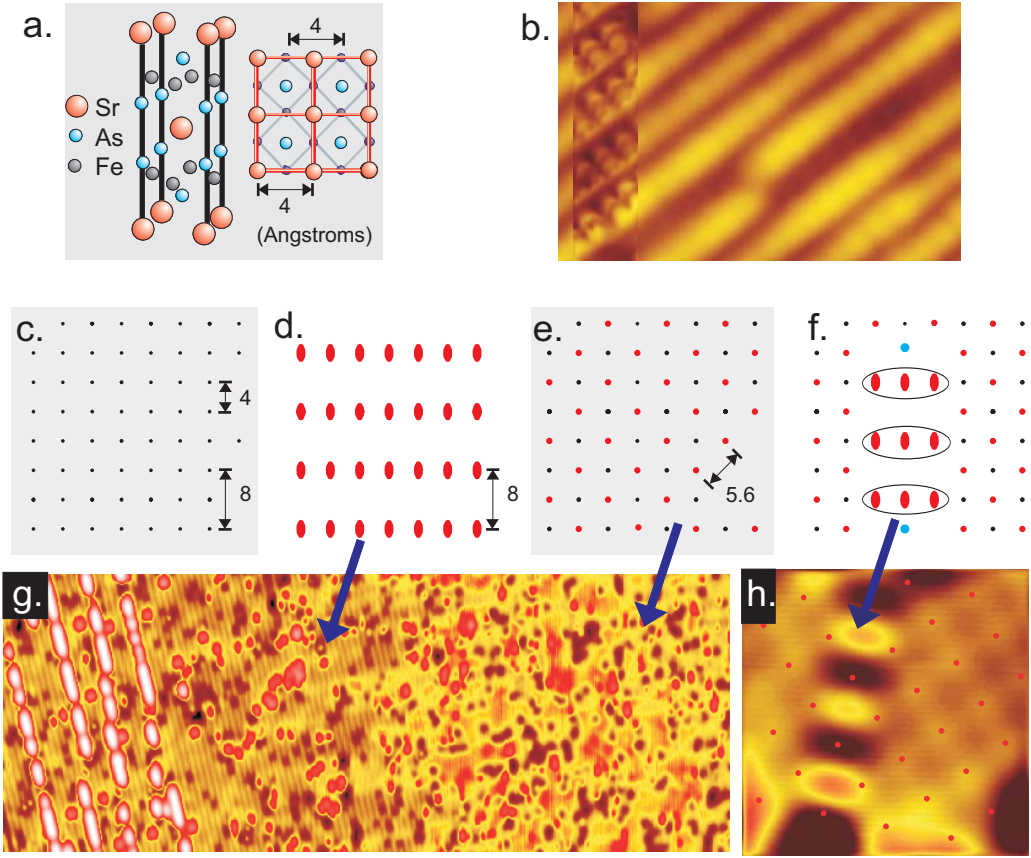

# Figure 3

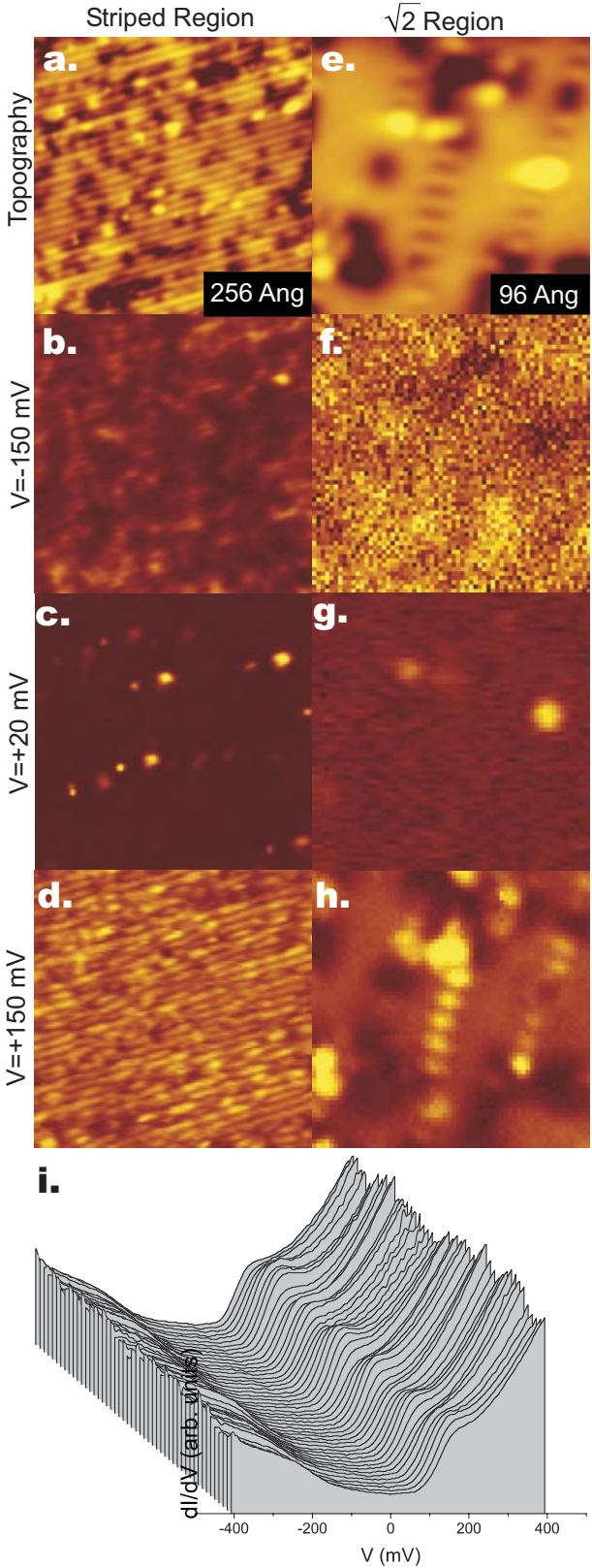

Figure 4

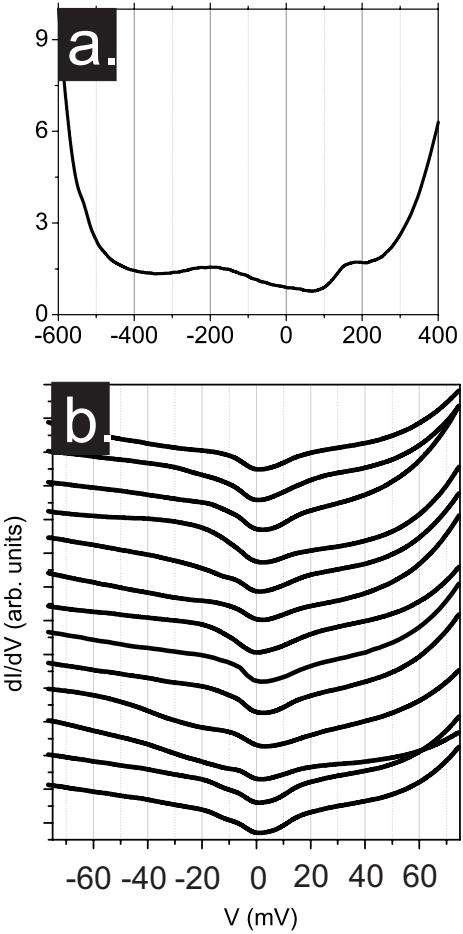